\documentclass[10pt,sigconf,letterpaper,nonacm]{acmart}

\usepackage[labelfont={small,bf},textfont={small,it}]{caption}
\usepackage{subcaption}
\usepackage{colortbl}
\usepackage{fancyref}
\usepackage{titlesec}
\usepackage{graphicx}
\usepackage{tabularx}
\usepackage[boxed]{algorithm2e}

\usepackage{mdframed}  
\usepackage{newfloat}
\usepackage{caption}
\DeclareFloatingEnvironment[placement={!ht}, name=Vignette]{vignette}  
\newcommand{\vignetteboxbegin}{ \begin{mdframed}[backgroundcolor=gray!10,shadow=true,shadowsize=0.5pt,shadowcolor=black!80,roundcorner=4pt] }
\newcommand{\vignetteboxend}{ \end{mdframed} \vskip -1em }

\def\BState{\State\hskip-\ALG@thistlm}

\titlespacing{\section}{0pt}{8pt}{4pt}
\titlespacing{\subsection}{0pt}{6pt}{3pt}
\titlespacing{\subsubsection}{0pt}{4pt}{2pt}
\titlespacing{\paragraph}{0pt}{4pt}{4pt}
\usepackage{xifthen}

\newcounter{mn}
\newcommand{\superscript}[1]{\ensuremath{{}^{\textrm{\scriptsize #1}}}}

\newcommand{\mntext}[1]{\colorbox{lime}{\begin{color}{black}#1\end{color}}}
\newcommand{\mn}[2][]{{\tiny\superscript{\mntext{\arabic{mn}}}}\marginpar{\scriptsize{
  \ifthenelse{\isempty{#1}}
  {\mntext{\parbox{0.95\marginparwidth}{\superscript{\arabic{mn}}~\raggedright{#2}}}}
  {\mntext{\parbox{0.95\marginparwidth}{\superscript{\arabic{mn}}#1 says~:~\raggedright{#2}}}}
}}\stepcounter{mn}}

\usepackage{relsize,etoolbox}
\AtBeginEnvironment{quote}{\smaller}

\newcommand{\KB}[1]{\textcolor{blue}{}}
\newcommand{\AH}[1]{\textcolor{purple}{[AH: #1]}}
\renewcommand{\AH}[1]{}
\newcommand{\PS}[1]{\textcolor{red}{}}
\newcommand{\eg}{{\it e.g.}}
\newcommand{\ie}{{\it i.e.}}

\renewcommand{\paragraph}[1]{\vspace*{0.05in}\noindent\textbf{#1}}

\title{Encryption without Centralization: Distributing DNS Queries Across Recursive Resolvers}
\renewcommand{\shorttitle}{Encryption without Centralization}

\author{Austin Hounsel}
\author{Paul Schmitt}
\affiliation{
    \institution{Princeton University}
    \country{United States}
}

\author{Kevin Borgolte}
\affiliation{
    \institution{Ruhr-University Bochum}
    \country{Germany}
}
\author{Nick Feamster}
\affiliation{
    \institution{University of Chicago}
    \country{United States}
}

\begin{document}
\begin{sloppypar}
\begin{abstract}

    Emerging protocols
    such as DNS-over-HTTPS (DoH) and DNS-over-TLS (DoT) improve the
    privacy of DNS queries and responses. While this trend
    towards encryption is positive, deployment of these protocols has in some cases
    resulted in further centralization of the DNS, which
    introduces new challenges. In particular, centralization has
    consequences for performance, privacy, and availability; a potentially
    greater concern is that it has become more difficult to control the choice
    of DNS recursive resolver, particularly for IoT devices.  Ultimately, the
    best strategy for selecting among one or more recursive resolvers
    may ultimately depend on circumstance, user, and even device.
    Accordingly, the DNS architecture must permit flexibility in allowing
    users, devices, and applications to specify these strategies. Towards this
    goal of increased de-centralization and improved flexibility, this paper
    presents the design and implementation of a refactored DNS resolver
    architecture that allows for de-centralized name resolution, preserving
    the benefits of encrypted DNS while satisfying other desirable properties,
    including performance and privacy.

\end{abstract}

%
%
\setcopyright{none}

\maketitle

\section{Introduction}

DNS has long been insecure and vulnerable to eavesdropping, but that reality is
changing, as protocols for encrypted DNS have recently been proposed and
deployed, notably DNS-over-TLS~(DoT) and DNS-over-HTTPS~(DoH). DoH, in
particular, has seen rapid adoption, as browser vendors have begun to move name
resolution functionality into applications themselves, whereas in the past it
was typically done at the OS level. DoH deployment depends on coordination
between the stub resolver at the client side (e.g., in the browser) and the
operator of the recursive resolver. In some cases, that
coordination is straightforward because the same organization operates both the
browser and the resolver (e.g., Google offers both a browser and a public DNS
service). In other cases, two organizations coordinate---as is the case where
Mozilla has collaborated with Cloudflare to deploy an encrypted DNS service in
Firefox, with Cloudflare serving as the primary recursive resolver.

DNS encryption is unquestionably a positive trend, but it is accompanied by a
troubling consequence: the {\em increased centralization of a critical part of
the Internet infrastructure}. This organizational centralization makes the DNS
infrastructure itself less resilient to disruption from misconfiguration,
attack, and outright manipulation. These threats are more than existential: An
attack on DNS infrastructure in 2016 rendered many websites
unreachable~\cite{nyt-mirai}. DNS queries are ripe for widespread
manipulation, resulting in information control and censorship. DNS
misconfiguration is also commonplace~\cite{borgolte2018cloud-strife}.
Centralization also has potentially adverse effects on competition,
introducing new barriers to entry as organizations who operate recursive
resolvers have access to DNS queries that can be used for a competitive
advantage in other market sectors, from content delivery to
advertising~\cite{borgolte2019dns}. The
increased centralization of DNS data into a handful of entities has also
raised privacy concerns about tracking users' browsing patterns through their
queries.

These high stakes have resulted in heated arguments and battles from
mailing lists to standardization bodies, such as the Internet Engineering Task
Force (IETF), whereby each of these stakeholders seeks to retain control over
the DNS. Faced with the prospect of losing visibility into DNS queries, some
ISPs have partnered with Mozilla to become trusted recursive resolvers.
Users who have privacy concerns over their ISPs eavesdropping on their DNS
traffic might be concerned by this development. Similarly, users who are more
concerned with advertisers seeing their browsing patterns would be rightfully
concerned that the IoT devices that they purchase from these same companies
default to sending DNS queries to the resolver of the same company (e.g., many of
Google's IoT products are hard-wired to use Google Public DNS as a
resolver~\cite{nest_dns}). Left behind in all of these power struggles is the user, who
often ends up relying on a centralized DNS operator based on
default configuration settings and the inertia that comes with changing
defaults.

Centralization trends continue: In 2017, more than 40\% of DNS traffic from
Tor was resolved via Google Public DNS. More recent statistics have shown that
more than 30\% of DNS queries to ccTLDs come from five large cloud providers,
two of whom offer their own centralized DNS service~\cite{moura2020clouding}. A
small number of organizations who operate DNS resolvers are gaining increased
market share. This centralization is occurring in spite of the fact that {\em
anyone} can operate a recursive resolver, and in fact hundreds of
organizations do just that. The trends towards centralization of this critical
part of the Internet are driven not by technical limitations, but rather by
ongoing trends of Internet consolidation, coupled with the bundling of critical
functionality like name resolution into applications themselves.

In this paper, we posit that {\em encryption of the DNS need not imply
centralization} of DNS queries at a resolver (or set of
resolvers) operated by a single organization and present an architecture that
permits the deployment of encrypted DNS protocols without coupling decisions
about resolvers to the default choices made by a particular browser vendor or other
connected device (e.g., consumer IoT devices). Specifically, we develop a
public, open-source, configurable stub resolver, based on {\tt dnscrypt-proxy}, that
allows users to configure how they want their encrypted DNS queries to be
distributed across a collection of resolvers. This custom proxy allows a user to
specify both the {\em set} of resolvers that any particular application or device
should use, as well as the {\em strategy} for how those queries should be
distributed across the set of resolvers that a user specifies. We have released
this stub resolver as an open-source fork of {\tt dnscrypt-proxy} so that others can
use and extend it.


\section{Background and Related Work}\label{sec:background}

In this section, we provide background on the development of encrypted DNS
protocols and explain how these protocols have led to a centralization of DNS.

\subsection{Encrypted DNS Protocols}

DNS queries and responses have historically been unencrypted, which has garnered
concern in recent years, given research that has demonstrated that
DNS traffic can be used to discover private information about users, ranging
from the websites and webpages that they visit to the ``smart'' devices that
they use (and how they operate them). 

T-DNS\cite{zhu2015connection} address security issues with DNS, such as lack of
confidentiality and amplified denial-of-service attacks. T-DNS has not been
widely adopted, but it served as the primary inspiration for DNS-over-TLS
(DoT)~\cite{rfc7858}. DNS-over-HTTPS (DoH)~\cite{rfc8484} aims to solve the same
problems as DoT, but uses HTTP as a transport protocol.  Other work investigated
the adoption of secure DNS and their real-world benefits.
Hounsel et al. measured web performance when using
encrypted DNS protocols and found that in some cases the newer protocols can
outperform conventional DNS~\cite{hounsel2020comparing}. Recent proposals from
Mozilla and Google involve sending DoH queries directly from the browser to a
recursive resolver~(sometimes simply referred to as a ``resolver'') as
configured in the browser (perhaps even by default, although as of this writing
the default settings have not yet changed). Similarly, the Android OS makes it
possible to route all DNS queries via DoT to a Google-operated
resolver~\cite{android-dot}.

\subsection{DNS Centralization}

From a user privacy perspective, DNS encryption is largely a positive
development, but an emergent side effect is the centralization of the protocol
and reduced local control. Clients that are configured to use DoT or DoH operate
using {\em centralized} architectures, whereby the client sends all DoT or DoH
queries to a single recursive resolver. Conventional DNS would initially appear
to share the same characteristics: a client typically sends all queries to its
local resolver, typically one that is configured via DHCP (\ie, configured by a
local network authority). Conversely, DoH has shifted name resolution
functionality {\it into applications themselves}, shifting control over
configuration to browser vendors, and in some cases, IoT device vendors. These
centralization trends have occurred rapidly, over a relatively short timespan.
In June 2018, Mozilla announced a partnership with Cloudflare to deploy DoH to
Firefox desktop users in the United States~\cite{mozilla_improving_dns_privacy}.
Mozilla implements DoH in the browser and Cloudflare operates a recursive
resolver that supports DoH. Initially, this option was enabled in Firefox
Nightly builds; over the course of 18 months, Mozilla transitioned to sending
all DNS queries to Cloudflare via DoH by default. In February 2020, Mozilla
enabled DoH by default for all Firefox users in the United States~\cite{mozilla_doh_rollout_2020}.

Foremski et al. find that the top 10\% of DNS recursors serve approximately 50\%
of DNS traffic~\cite{foremski2019dns-observatory}.  Moura et
al.~\cite{moura2020clouding} also encounter centralization in their study of DNS
requests to two country code top-level domains (ccTLD), with five large cloud
providers being responsible for over 30\% of all queries for the ccTLDs of the
Netherlands and New Zealand.  
Recent developments suggest that these trends could be reversed; for example,
Hoang et al.~\cite{hoang2020kresolver} propose and evaluate K-resolver, which
distributes queries over multiple DoH recursors in Firefox, so that no single
resolver can build a complete profile of the user and each recursor only learns
a subset of domains the user resolved. Arkko et al. propose several strategies
for distributing DNS queries and discuss the performance and privacy trade-offs
of each strategy~\cite{arkko}. Other previous work also shows that
distributing DNS queries across multiple resolvers in various fashions can yield
acceptable performance~\cite{hounsel2020comparing,boettger2019empirical}. This
paper extends this past work, adding additional distribution strategies. Our
work also shifts control of name resolution decisions out of individual
applications and allows all devices passing through our proxy to benefit from
DNS query distribution strategies. This is critical as many devices,
particularly IoT devices, make DNS configuration opaque and challenging. Our
design also allows for rule-based DNS strategy selection such as matching on
client MAC or IP addresses.

\section{System Design and Implementation}\label{sec:stub_design}
\textbf{Decisions about DNS resolution should occur at a single place: a
separate stub resolver that performs resolution for all applications and
devices for which a user or users have a common set of preferences.} This has 
traditionally been the role of an operating system stub resolver running on a host device or a router, but in recent years, applications have performed DNS resolution on their own.
By returning to how DNS resolution is traditionally performed, the stub resolver is able to provide applications, ISPs, and users a single place to define how DNS name resolution should occur.
Such modularization also enables us to experiment with new features for DNS resolution.

\subsection{Overview}
We propose that a stub resolver perform DNS resolution as follows:
\begin{enumerate}
    \item the stub resolver discovers a collection of upstream resolvers that
        support DoH, along with various characteristics of those resolvers
        (\eg, geographic location). This configuration can manually performed by the device owner, or automatically through negotiation with an upstream network operator (\eg, via DHCP).
    \item a user can specify specific requirements or preferences about preferred
        resolvers or goals (\eg, a preference to avoid a specific location, geography
        or ISP; or a preference of privacy over performance or vice versa); or,
        alternatively, an explicit selection; 
    \item the stub selects DoH resolvers by matching availability with user preference; and 
    \item the stub distributes queries across multiple DoH resolvers to reduce
        centralization based on a user-specified strategy. 
\end{enumerate}
\noindent
This design bears some resemblance to the behavior of
an operating system stub resolver, but bears the additional characteristics of
configurability, and the option to place the stub resolver at a point in the
network that is independent of any device but common to a set of user (or
users) who share common preferences, such as a home network router.

For wide-scale deployment, we envision that operating systems will implement
the proposed stub resolver, similar to Windows and Android adding native
support for encrypted DNS~\cite{windows-doh,android-dot}.  We
acknowledge that this solution can be circumvented: Applications and IoT
devices could bypass the proxy by directly querying resolvers of their choice,
especially if DNS queries are encrypted.

\subsection{Design Principles}\label{sec:stub}
A separate stub resolver that can resolve queries for {\em all} users and
devices who share preferences about performance, privacy, security, and other
considerations is an appropriate location to address DNS centralization.

\paragraph{The stub resolver should not presume an outcome with respect to the
set of resolvers or the strategies for distributing queries across them.}
As described later in this section, we envision a stub resolver that affords
many possible configuration options, and \Fref{sec:distribution_strategies} explores
one such customization option that involves decentralizing DNS queries across
multiple recursive resolvers, using one of many possible distribution
strategies. Such a level of configurability required only modest modifications
to existing DNS stub resolvers, as we describe in
Section~\ref{sec:implementation}.

\paragraph{Users should be able to choose how DNS queries are resolved, to
implement these choices for all devices on their network.} In contrast to the
status quo, where browsers perform encrypted DNS resolution on behalf of users
and where other
Internet-connected devices may select their own DNS resolution
mechanism---resolving {\em all} DNS queries in a separate stub resolver that a
user can configure and customize provides more choice to the user. (A separate,
important question concerns whether users understand the consequences of these
choices, and how to make those choices visible. We are currently conducting a
user survey to understand this question.)

\paragraph{All stakeholders should be able access this point of control, to
allow for optimizations and customizations.} In contrast
to the current architecture, where browser and device vendors hold control over
which entities can be recursive resolvers and which resolvers are
selected by default, a separate stub resolver can potentially be controlled by
{\em any} of the stakeholders. Naturally, we expect that there will be push and
pull, and even cooperation (or collusion) among these entities. But,
modularizing the DNS resolution process in this fashion will make those actions
{\em visible}: An anti-competitive maneuver such as restricting an API to
configure the stub, or collusion between content providers and browser vendors
would be plainly apparent in such an architecture---and likely reversible, if
not through alternate implementations, then via regulatory mechanisms.

\subsection{Prototype Implementation}\label{sec:implementation}

We forked the open-source {\tt dnscrypt-proxy} stub resolver~\cite{dnscrypt-proxy} to support new strategies and policies to de-centralize DNS queries (\Fref{sec:distribution_strategies}).
We extended the \texttt{getOne()} function within \texttt{serversInfo.go}, which indexes into an array of upstream resolvers for each query based on which distribution strategy is specified in the configuration file~\cite{dnscrypt-proxy_serversInfo}.
The prototype is publicly available 
(\url{https://github.com/noise-lab/ddns}).
We include instructions for installing and running the proxy, as well as code for running performance measurements.
The proxy supports DoH and DNSCrypt, and it can run on both host devices and routers.
In the configuration file for the proxy, users can specify which strategy they wish to use and which resolvers they wish to distribute queries over.

There are several possible applications for distributing queries between multiple resolvers for a device or local network.
The main use case we envision is giving users control over the share of their DNS queries that various resolvers collect, which may enhance their privacy.
We also envision giving network operators additional control over how encrypted DNS resolution is performed by devices they own on their networks.
For example, network operators in an enterprise environment may want to map queries from certain devices to (encrypted) DNS resolvers that they operate, while allowing all other devices to use other resolvers. 
This would enable network operators to support split-horizon DNS while also balancing concerns for DNS privacy.
Home network operators may also wish to forward queries from certain devices to certain resolvers to limit the information certain parties have about them.
IoT devices like Google Nest may communicate with Google servers for functionality, so it may make sense to forward DNS queries for these devices to resolvers not owned by Google.

Although we do not advocate for a particular query distribution strategy, we
argue that the proxy should make a default choice to achieve wide-scale
deployment, as existing applications such as browsers already do.
Future work should conduct user studies to inform what the default
configuration should be.

\subsection{Distributing DNS Queries}\label{sec:distribution_strategies}

In this section, we describe several strategies for DNS resolution that the
prototype implements.

\paragraph{Hash-Based Distribution}
In a hash-based distribution, second-level domain names (SLDs) are hashed to index into a list of resolvers, meaning that queries for the same SLD will always be sent to the same
resolver.
For example, all queries issued by a client for \texttt{google.com} and \texttt{images.google.com} will be sent 
to the same resolver.
Furthermore, if the same client later queries \texttt{images.google.com}, the query will be forwarded to the same resolver as before.
This strategy ensures that no two resolvers receive queries for the same domain name, but some resolvers may receive a larger share of queries.
Furthermore, this strategy may be less robust to failure:
If a resolver fails, users may not be able to perform DNS resolution for
certain domain names.

\paragraph{Random Distribution} In the random distribution strategy, queries
are randomly sent to a set of defined resolvers R, resulting in each resolver
handling $ \frac{1}{R} $ of the client's queries.  This is a simple
strategy, and recovery from failure is simple: If a resolver is down, users can 
send their queries to another random resolver. We re-use the random
distribution code that was originally implemented by {\tt dnscrypt-proxy}.

\paragraph{Round-Robin Distribution}
Using this strategy, queries are sequentially striped across a set of resolvers
R. The round-robin strategy results in each resolver would be assigned $
\frac{1}{R} $ of the client's queries.
This strategy ensures that queries are evenly distributed over multiple resolvers, but it enables multiple resolvers to receive queries for the same domain name over time.
As with random distribution, round-robin distribution may provide users with
more resilience to failure.

\section{Prototype Evaluation}\label{sec:case-study}

We explore how the architecture can enable de-centralization of queries,
and evaluate its effect on CDN localization, performance, and privacy.

\begin{figure}[t]
    \includegraphics[width=0.75\linewidth]{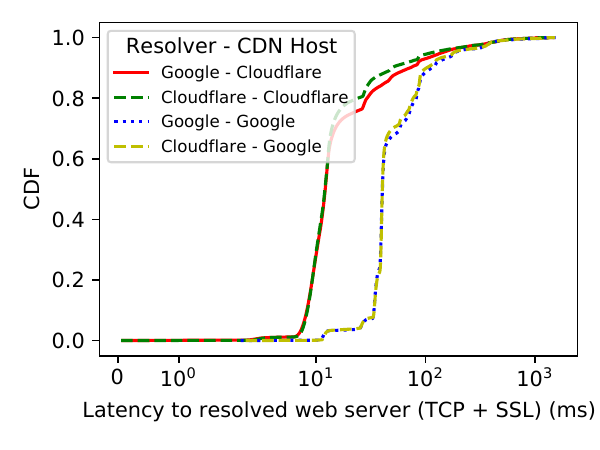}
    \caption{TCP and SSL setup times to CDN servers operated by Cloudflare and Google. Each line shows setup times when a particular DNS resolver is used for either Cloudflare or Google hosted content.}
    \label{fig:cdn_localization}
\end{figure}

\subsection{Performance Effects}

\paragraph{How Does Query Distribution Affect CDN Localization?}
We seek to understand whether distributing queries across multiple
recursive resolvers could negatively affect CDN localization.
We fetch
each HTTP request from the \textit{requests\_desktop} table in the HTTP Archive
for October 2020~\cite{http_archive}. We also use information provided by the
HTTP Archive to determine which CDN each domain name hosts its content on, if
applicable. For comparison purposes, we study content that is hosted by
Cloudflare or Google, as these providers also both operate DNS resolvers. We
resolve the domain names twice, once using Cloudflare's DNS and once using
Google's, for each request that was hosted by either Cloudflare or Google's CDN
networks, and measure the latency for TCP and SSL connection setup to the
resolved IPs from a 500 Mbps residential fiber connection. 

Figure~\ref{fig:cdn_localization} shows the 
cumulative distribution function (CDF) for combined TCP and TLS
setup times for a given resolver and CDN. For example, the line that corresponds
to "Google - Cloudflare" shows combined TCP and TLS setup times
when Google's resolver is used to resolve the domain names of content hosted on Cloudflare's CDN. We find
that concerns over whether distributing queries over multiple resolvers will
affect CDN localization are not significant in our experiment. When either
Google's resolver or Cloudflare's resolver is used to resolver Google-hosted
content, TCP and TLS setup times follow the same distribution. The distributions
for each resolver are slightly different when Cloudflare content is resolved,
but for the most part, the distributions are very similar. We note that Google
and Cloudflare host two of the most popular resolvers, but we
expect similar results with any resolver that is widely distributed.

\paragraph{What is the Effect of Query Distribution on Page Load Times?}
We performed page loads from several vantage points for 20 days.
We created an Amazon EC2 instance at four vantage points--Ohio, North Virginia, California, and Oregon--that each ran Debian Linux.
To perform our measurements, we extended a Docker image created by Hounsel et al. that performs page loads using a headless version of Mozilla Firefox 84.0.1 controlled by Selenium~\cite{hounsel2020comparing}.
Each page load is performed within a separate Docker container.
Once we launch a container, we first run our fork of {\tt dnscrypt-proxy} within the container with a configuration file that corresponds to the strategy that we intend to measure.
We then modify \texttt{/etc/resolv.conf} to use our stub resolver, and we initiate a page load.
Once the page load completes, an HTTP Archive Object (HAR) corresponding to the page load is extracted from the container, and we close the container.
We read the timing for the \texttt{onLoad} event in each HAR to measure page
load times.~\footnote{We disabled the DNS cache for {\tt dnscrypt-proxy}. Firefox
maintains its own in-memory DNS cache, but because each page load was
performed within a separate Docker container and because Firefox clears its
cache upon exit each page load uses clean DNS and HTTP caches .}
We used the top 1,000 websites on the Tranco top-list for December 12th, 2020 to perform measurements~\cite{le2019tranco}.

\begin{figure}[t!]
    \centering
    \begin{subfigure}[t]{0.4\textwidth}
        \centering
        \includegraphics[width=\textwidth]{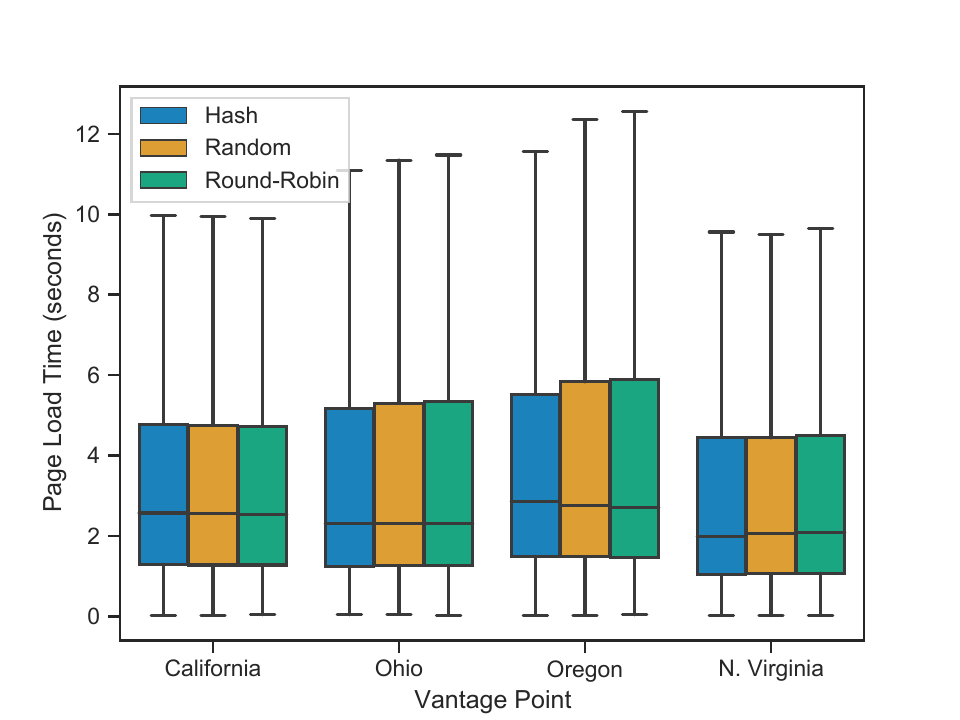}
        \caption{Query distribution strategies.}
        \label{fig:pageload_boxplots_multiresolver}
    \end{subfigure}
    \hfill
    \begin{subfigure}[t]{0.4\textwidth}
        \centering
        \includegraphics[width=\textwidth]{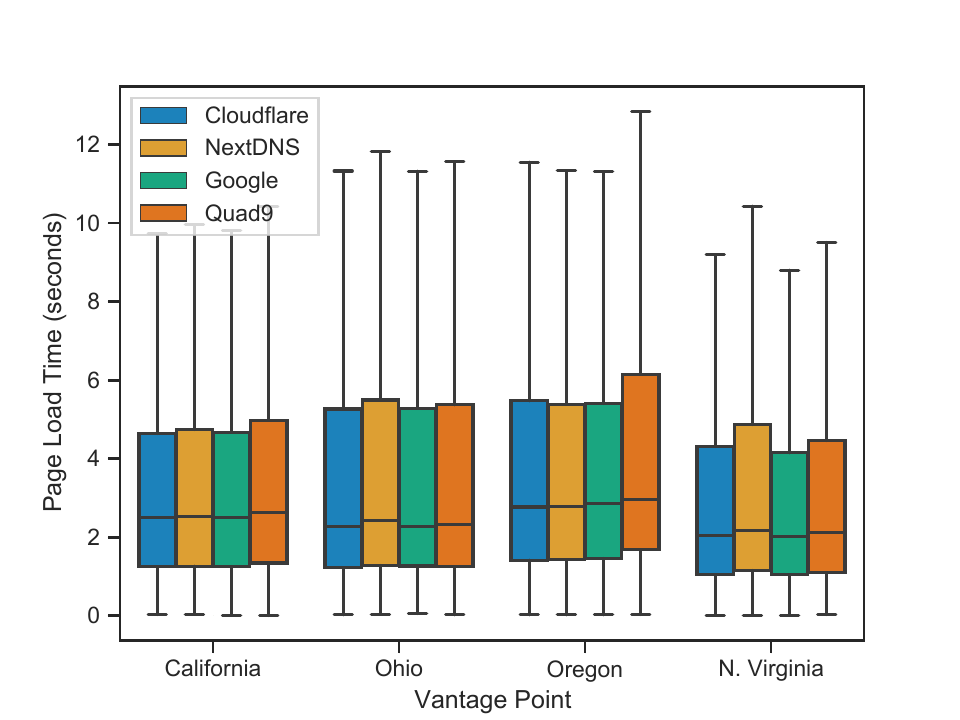}
        \caption{Individual resolvers.}
        \label{fig:pageload_boxplots_singleresolver}
    \end{subfigure}
    \caption{Page load times from each vantage point using query distribution models and individual resolvers.}
    \label{fig:pageload_boxplots}
\end{figure}


\begin{table}[t]
  \small
  \centering
  \begin{tabularx}{\columnwidth}{Xrrrr}
    \rowcolor{white}
      \toprule
      \textbf{Location}
      & \textbf{Hash}
      & \textbf{Random}
      & \textbf{Round-robin}
      \\
      \midrule
        California
        & 2.57
        & 2.55
        & 2.52
        \\
        Ohio
        & 2.30
        & 2.30
        & 2.31
        \\
        Oregon
        & 2.85
        & 2.75
        & 2.70
        \\
        N. Virginia
        & 1.97
        & 2.06
        & 2.08
        \\
        \bottomrule
  \end{tabularx}
  \caption{Median page load times (in seconds) from each vantage point using each query distribution strategy.}
  \label{tab:pageload_medians_multiresolver}
\end{table}


\begin{table}[t]
  \small
  \centering
  \begin{tabularx}{\columnwidth}{Xrrrr}
    \rowcolor{white}
      \toprule
      \textbf{Location}
      & \textbf{Cloudflare}
      & \textbf{NextDNS}
      & \textbf{Google}
      & \textbf{Quad9}
      \\
      \midrule
        California
        & 2.51 
        & 2.53
        & 2.50
        & 2.63
        \\
        Ohio
        & 2.28
        & 2.43
        & 2.27
        & 2.32
        \\
        Oregon
        & 2.77
        & 2.78
        & 2.85
        & 2.96
        \\
        N. Virginia
        & 2.05
        & 2.18
        & 2.03
        & 2.12
        \\
        \bottomrule
  \end{tabularx}
  \caption{Median page load times (in seconds) from each vantage point using a single resolver for all DNS queries.}
  \label{tab:pageload_medians_singleresolver}
\end{table}

For our query distribution strategies, we used the DoH resolvers provided by Cloudflare, Google, Quad9, and NextDNS.
We chose these resolvers due to their popularity and their support in major browsers.
For example, as of January 13th, 2021, Cloudflare and NextDNS are the two default DoH providers that are listed in Mozilla Firefox 84.0.1.
Similarly, Google Chrome automatically upgrades users of Cloudflare, NextDNS, Google, and Quad9's resolvers to DoH~\cite{chrome_doh_upgrade_blog,chrome_doh_upgrade_providers}.
In addition to performing pageloads with the query distribution strategies using these resolvers, we also measure page load times when using each of these resolvers on their own for all DNS queries.

Figure~\ref{fig:pageload_boxplots_multiresolver} shows page load times for each query distribution strategy, and Figure~\ref{fig:pageload_boxplots_singleresolver} shows page load times for each resolver.
Table~\ref{tab:pageload_medians_multiresolver}
and Table~\ref{tab:pageload_medians_singleresolver} show median page load times for
each strategy and resolver.
First, for most vantage points, each strategy performs similarly in terms of
median page load times, although
the largest gap in performance was in Oregon between the hash strategy and round-robin strategy, with the hash model performing 150 ms slower.
The largest difference between two strategies was lower in other vantage points, with 50 ms in California, 10 ms in Ohio, and 110 ms in N. Virginia.
Page load times are similar with each resolver, although Quad9 does perform slower in Oregon.

\subsection{Privacy}

\paragraph{How Does Query Distribution Affect Domain Names Seen By Resolvers?}
We next study how many unique domain names are seen by each DNS resolver over
time if different query distribution strategies are used.
To do so, we use a real-world dataset of anonymized DNS queries for approximately 100 homes connected to a fiber-to-the-home (FTTH) network in a residential neighborhood in Cleveland, OH~\cite{allman2020putting}.
Each home is connected by a gateway device that proves a single public IP address for each home through NAT.
This dataset consists of queries issued over a seven-day period each month during 2018 (\ie, 12 weeks).
Full details about the dataset and collection methodology can be found in Allman et al~\cite{allman2020putting}.
We first group the queries that each source IP address (\ie, each home) issued together.
We then extract the timestamp for each query, ordering each address' queries
by the time in which they were issued.  Finally, we simulate each of our query
distribution strategies "after the fact" on each address' ordered list of
queries.

Previous work has observed the random and round-robin distribution strategies
may decrease user privacy in the long term, as all resolvers learn more domain
names~\cite{arkko}.
Our work (and system) provides a way to quantify the effects of both these
strategies and other alternatives that may be designed in the future.
For example, other strategies may be more beneficial for privacy (\eg, hash-based strategies); we believe it is useful to provide points of comparison.

When the hash strategy is used, each resolver sees fewer unique domain names for each address than when the other strategies are used.
When four resolvers are used for the hash strategy, the strategy stabilizes with an average of $\approx$25\% of unique domain names for each address seen by each resolver.
On the other hand, when four resolvers are used for the random strategy and the round-robin strategy, the strategies stabilize with an average of $\approx$50\% of queries seen by each resolver.
Interestingly, after just one week, the random and round-robin strategies stabilize with a mean of $\approx$45\% of unique domain names seen by all four resolvers, compared to $\approx$50\% after 12 weeks of data.
We note that with the round-robin and random distributions, every resolver will quickly see popular domain names (\eg, \texttt{google.com} and \texttt{facebook.com}), but they won't each see the domain names that were queried a small number of times.
Over time, each resolver may see the same domain names, but they may not see this data quickly.

\balance\section{Conclusion}

This paper has argued for a re-decentralization of the DNS.  
Users may prefer one distribution strategy over another.  In this
vein, we believe that this paper lays the groundwork for much future work in
both research and industry, as we explore various alternative strategies for
resolving and distributing encrypted DNS queries. This paper provides one such
starting point as a proof-of-concept.

\paragraph{Acknowledgments.} This work was funded in part by NSF Award
CNS-1953513.

\pagebreak

\bibliographystyle{ACM-Reference-Format}
\bibliography{paper,proceedings}
\end{sloppypar}

\end{document}